\documentclass[a4paper,cits]{PoS}
\pdfoutput=1
\usepackage{amsmath,bbm,slashed,dsfont,MnSymbol}

\newcommand{\be}{\begin{equation}}
\newcommand{\ee}{\end{equation}}
\newcommand{\Z}{\mathcal{Z}}
\newcommand{\D}{\mathcal{D}}
\newcommand{\expv}[1]{\left \langle #1 \right \rangle}
\newcommand{\M}{\mathcal{M}}
\newcommand{\Dp}{\slashed{D}(\mu_I)}
\newcommand{\Dm}{\slashed{D}(-\mu_I)}
\newcommand{\tr}{\textmd{tr}}
\renewcommand{\d}{\textmd{d}}
\newcommand{\light}{{ud}}
\renewcommand{\O}{\mathcal{O}}

\title{QCD phase diagram with isospin chemical potential\thanks{This
contribution contains the combined proceedings of the two talks
presented by the authors at the conference: 
`QCD with isospin chemical potential: low densities and Taylor expansion'
and 
`QCD with isospin chemical potential: pion condensation'.}}

\ShortTitle{QCD phase diagram with isospin chemical potential}

\author{Bastian B. Brandt and Gergely Endr\H{o}di
         \\
        Institute for Theoretical Physics, Goethe University,
        Max-von-Laue-Strasse 1, 60438 Frankfurt am Main, Germany \\
        E-mail: \email{brandt@th.physik.uni-frankfurt.de},
        \email{endrodi@th.physik.uni-frankfurt.de}}

\abstract{
In this contribution we investigate the phase diagram of QCD in the presence of an 
isospin chemical potential. To alleviate the infrared problems of the theory 
associated with pion condensation, we introduce the pionic source as an
infrared regulator. We discuss various methods to extrapolate the results to
vanishing pionic source, including a novel method based on the singular value
spectrum of the massive Dirac operator, a leading-order reweighting and a spline
Monte-Carlo fit. Our main results concern the phase transition boundary between
the normal and the pion condensation phases and the chiral/deconfinement 
transition temperature as a function of the chemical potential. In addition, we perform 
a quantitative comparison between our direct results and a Taylor-expansion obtained 
at zero chemical potential to assess the applicability range of the latter.
}

\FullConference{34th annual ``International Symposium on Lattice Field
                 Theory''\\
                 24-30 July 2016\\
                 University of Southampton, UK}

\begin{document}

\section{Introduction}

Quantum Chromodynamics (QCD) is the theory of the strong interactions. It
describes how protons and neutrons are built up of elementary particles: quarks
and gluons. For various physical applications ranging from the evolution of the
early universe through neutron star physics to heavy-ion collisions, it is of
interest how quarks and gluons behave if the system is heated up or is
compressed. The relevant parameters in this context are the temperature $T$ and
the quark densities $n_f$ for each quark flavor $f$. For most of the above
scenarios only the light quarks $f=u,d,s$ contribute. In the grand canonical 
ensemble, the densities are traded for the corresponding chemical potentials
$\mu_f$ as conjugate parameters. Instead of working in the flavor basis, it is
customary to introduce the three independent combinations
\be
\label{eq:chpot}
\mu_B = \frac{3}{2}(\mu_u+\mu_d), \quad\quad \mu_I=\frac{1}{2}(\mu_u-\mu_d),
\quad\quad
\mu_S=-\mu_s\,,
\ee
being the baryon, isospin, and strangeness chemical potentials. On timescales
relevant for the strong interactions -- where flavor-changing weak processes are
ineffective -- all three densities are conserved and working with the
corresponding chemical potentials is justified.

While $\mu_B$ and $\mu_S$ couple to baryon number and strangeness, $\mu_I$
couples to the difference of the number of protons and neutrons. Thus, it is a
relevant parameter for systems with an asymmetry between protons and neutrons.
Two prime examples that exhibit such an asymmetry are the core of neutron stars
and the initial state of heavy-ion collisions. Since neutrons dominate over 
protons in both cases, these systems are characterized by a negative isospin
chemical potential $\mu_I<0$. At the same time, the nonzero baryon number
implies $\mu_B>0$.

One of the most effective systematic approaches to study the physics of quarks 
and gluons is by means of numerical simulations of the QCD path integral on a
Euclidean space-time lattice. These simulations employ standard importance
sampling techniques, which rely on the probabilistic interpretation of 
$\exp(-S_{\rm QCD})$, where $S_{\rm QCD}$ is the QCD action. However, for
nonzero baryon or strangeness chemical potentials, $S_{\rm QCD}$ becomes
complex. This is the so-called complex action problem (sign problem) that 
invalidates direct Monte-Carlo simulations. Fortunately, the sign problem does
not affect nonzero isospin, since $S_{\rm QCD}$ remains real for $\mu_I\neq 0$. 

In the present contribution we consider the impact of a nonzero isospin chemical
potential on the phase structure of QCD, and set $\mu_B=\mu_S=0$. Although not 
directly related to the above described physical situations, this system
captures an interesting phenomenon that might be relevant for neutron stars and
for nuclear physics -- the condensation of
pions~\cite{Migdal:1978az,Ruck:1976zt}. Indeed, it is known from chiral
perturbation theory~\cite{Son:2000xc} that at the threshold value
$\mu_I=m_\pi/2$, a second-order phase transition takes place and the ground
state transforms to a Bose-Einstein condensate of pions.

Besides this phenomenological motivation, there are various conceptual and
technical similarities between the theory at nonzero isospin and that at nonzero
baryon density. In both cases the zero-temperature behavior of the theory
involves the so-called Silver Blaze phenomenon~\cite{Cohen:2003kd} -- where 
the chemical potential affects the fermionic action but has no impact on the 
ground state -- followed by particle creation beyond a threshold chemical 
potential. In addition, beyond the threshold a proliferation of near-zero 
eigenvalues takes place, leading to an ill-conditioned fermion matrix and 
numerical problems for its inversion, i.e., for the simulation algorithm.
This necessitates the use of an infrared regulator that we denote by $\lambda$
below. Understanding these concepts and facing these technical challenges 
in the (sign-problem-free) $\mu_I\neq0$ theory may give us insight on how to
assess the $\mu_B\neq0$, $\mu_S\neq0$ system in the future, once the sign
problem has been circumvented.

Lattice QCD with nonzero $\mu_I$ has already attracted considerable amount of
interest, see, e.g.,
Refs.~\cite{Kogut:2002tm,Kogut:2002zg,Kogut:2004zg,deForcrand:2007uz,
Detmold:2012wc,Endrodi:2014lja}. In this contribution we improve our
understanding of this system by simulating at the physical value of the pion
mass and by using an improved staggered action. In addition, we develop a novel
method for the extrapolation of the infrared regulator $\lambda$ to zero. This 
method involves the singular values of the massive Dirac operator, which 
we discuss for the first time on the lattice in this context. The extrapolated
results allow for a determination of the phase boundary between the normal and
the pion condensed phase and of the chiral/deconfinement transition line in the
phase diagram. Yet another use of our results is the possibility to directly
check the applicability range of a Taylor-expansion in $\mu_I$ around
$\mu_I=0$. Our quantitative comparison gives a hint on how far similar
expansions in the baryon chemical potentials might be reliable.

\section{Simulation setup and observables}
\label{sec:setup}

We consider three-flavor QCD with degenerate light quark masses $m_u=m_d\equiv
m_{\light}$ and strange quark mass $m_s$ at temperature $T$ and in a finite
volume $V$. The theory is discretized on a $N_s^3\times N_t$ space-time lattice
with spacing $a$ using stout smeared rooted staggered quarks and tree-level
Symanzik improved gluons. The quark masses are tuned to their physical values
along the line of constant physics $m_f(\beta)$. This tuning, together with the
simulation algorithm and the action at $\mu_I=0$ is detailed in
Ref.~\cite{Borsanyi:2010cj}, while the implementation of the chemical potential
follows Ref.~\cite{Endrodi:2014lja}. The partition function of the system is
given in terms of the path integral over the gluon links $U$,
\be
\Z = \int \D U \, e^{-\beta S_G}\, (\det \M_\light)^{1/4}\,(\det \M_s)^{1/4} \,,
\label{eq:Z}
\ee
where $\beta=6/g^2$ denotes the inverse gauge coupling, $S_G$ is the gluon
action, $\M_\light$ is the light quark matrix in up-down basis and $\M_s$ is the
strange quark matrix,
\be
\M_\light = \slashed{D}(\tau_3 \mu_I) + m_\light \mathds{1} + i\lambda \eta_5
\tau_2 = 
\begin{pmatrix}
 \Dp+m_\light & \lambda \eta_5 \\
 -\lambda \eta_5 & \Dm + m_\light
\end{pmatrix}, \quad\quad
\M_s = \slashed{D}(0) + m_s\,.
\label{eq:M}
\ee
Here, $\slashed{D}$ is the massless Dirac operator and $\tau_i$ denote the Pauli
matrices. The off-diagonal term in $\M_\light$ involves the pionic source
$\lambda$, which was briefly mentioned in the introduction and whose role will
be explained in more detail below. The matrix $\eta_5=(-1)^{n_x+n_y+n_z+n_t}$ is
positive (negative) on even (odd) sites and is the staggered equivalent of
$\gamma_5$. A direct Monte-Carlo simulation of this system is feasible because
of positivity for $\lambda>0$ and $m_s>0$,
\be
\det \M_\light = \det \left( |\Dp+m_\light|^2 + \lambda^2 \right) > 0,
\quad\quad \det \M_s = \det \left( | \slashed{D}(0) + m_s|^2 \right)^{1/2} >
0\,,
\label{eq:positive}
\ee
which follows from the hermiticity relations
\be
\eta_5\tau_1 \M_\light \tau_1 \eta_5 = \M_\light^\dagger, \quad\quad
\eta_5 \M_s \eta_5 = \M_s^\dagger\,,
\ee
that are satisfied due to chirality $\{\slashed{D},\eta_5\}=0$ and the relation
$\slashed{D}^\dagger(\mu_I) = -\Dm$. In the simulation algorithm the
representations with the squared operators $|\slashed{D}+m|^2$ are used (for
the strange quark, an even-odd block diagonalization enables us to get rid of
the square root in Eq.~(\ref{eq:positive})).

It is instructive to discuss the flavor symmetries of $\M_\light$. Besides the 
anomalous and baryonic $\mathrm{U}(1)$ symmetries, at $\mu_I=\lambda=m_\light=0$
it possesses an $\mathrm{SU}_L(2)\times \mathrm{SU}_R(2)$ chiral symmetry. This
is broken down to $\mathrm{SU}_V(2)$ by the light quark mass and, subsequently,
to $\mathrm{U}_{\tau_3}(1)$ by the chemical potential. The subscript in the
latter case indicates that the generator of the remaining symmetry 
is $\tau_3$. This $\mathrm{U}(1)$ symmetry is spontaneously broken by any of the
expectation values $\expv{\bar\psi \eta_5 \tau_1 \psi}$, $\expv{\bar\psi \eta_5
\tau_2 \psi}$, signaling pion condensation. The spontaneous breaking of this
continuous symmetry implies the presence of a Goldstone mode that leads to
infrared problems for the numerical algorithm. Our implementation
in~(\ref{eq:M}) corresponds to an explicit breaking that selects the $\tau_2$
direction for the breaking and makes the would-be massless mode a
pseudo-Goldstone boson. This small explicit breaking enables us to observe 
spontaneous symmetry breaking by looking at the expectation value
$\expv{\bar\psi \gamma_5 \tau_2 \psi}$ and, at the same time, alleviates the
infrared problem mentioned above. While the simulations are performed at
$\lambda>0$, at the end of the analysis we need to take the physical limit
$\lambda\to0$ by means of an extrapolation.

The observables we consider are, besides the already mentioned pion condensate,
the light quark condensate and the isospin density,
\be
\expv{\pi} = \frac{T}{V}\frac{\partial \log\Z}{\partial \lambda}, \quad\quad
\expv{\bar\psi\psi} = \frac{T}{V}\frac{\partial \log\Z}{\partial m_\light},
\quad\quad
\expv{n_I} = \frac{T}{V}\frac{\partial \log\Z}{\partial \mu_I}\,.
\label{eq:pbpdef}
\ee
Using Eq.~(\ref{eq:Z}) and the relations in Eq.~(\ref{eq:positive}), we
specifically have
\be
\begin{split}
\expv{\pi} &= \frac{T}{2V} \expv{\tr
\frac{\lambda}{|\Dp+m_\light|^2+\lambda^2}}, \\
\expv{\bar\psi\psi} &= \frac{T}{2V} \expv{\textmd{Re } \tr
\frac{\Dp+m_\light}{|\Dp+m_\light|^2+\lambda^2}}, \\
\expv{n_I} &= \frac{T}{2V} \expv{ \textmd{Re }\tr \frac{(\Dp+m_\light)^\dagger
\Dp'}{|\Dp+m_\light|^2+\lambda^2}}, \\
\end{split}
\label{eq:traces1}
\ee
where the prime denotes the differentiation of the operator with respect to
$\mu_I$. While the isospin density is free of ultraviolet divergences, the two
condensates are subject to renormalization. For $\expv{\bar\psi\psi}$, additive
divergences appear for nonzero mass, whereas the analogous situation occurs for
$\expv{\pi}$ for $\lambda>0$. On the one hand -- since we are working at fixed
physical quark mass -- additive renormalization is necessary for $\bar\psi\psi$,
which can be achieved by subtracting its value at $T=\mu_I=0$. On the other
hand, in the $\lambda\to0$ limit no such procedure is required for the pion
condensate.

In addition, multiplicative divergences also appear in both condensates, which
can be canceled by the corresponding multiplicative renormalization factors
$Z_\pi$ and $Z_{\bar\psi\psi}$. From the definitions~(\ref{eq:pbpdef}) it is
clear that $Z_\pi=Z_\lambda^{-1}$ and $Z_{\bar\psi\psi}=Z_{m_{\light}}^{-1}$. At
first sight it may appear that the two renormalization constants are
independent, but in fact they coincide in renormalization schemes independent of
$\mu_I$. To see this, notice that at $\mu_I=0$ the mass and $\lambda$ are
related by an isospin rotation and are thus equivalent -- consequently,
$Z_{m_{\light}}=Z_{\lambda}$ holds. Furthermore, neither of these
renormalization constants are affected by the chemical potential, thus the
equality holds for arbitrary $\mu_I$. Instead of calculating $Z_{m_{\light}}$, 
we multiply the condensates by $m_{\light}$ to obtain a combination in which 
the renormalization constants cancel. Altogether, the renormalized observables
read
\be
\label{eq:renobs}
\Sigma_{\bar\psi\psi} = \frac{m_{\light}}{m_\pi^2 f_\pi^2}
\left[ \expv{\bar\psi\psi}_{T,\mu_I} - \expv{\bar\psi\psi}_{0,0} \right] + 1,
\quad\quad
\Sigma_{\pi} = \frac{m_{\light}}{m_\pi^2 f_\pi^2} \expv{\pi}\,,
\ee
where we also included a normalization factor involving the pion mass
$m_\pi=135 \textmd{ MeV}$ and the chiral limit of the pion decay constant
$f_\pi=86 \textmd{ MeV}$ for convenience. In this normalization,
zero-temperature leading-order chiral perturbation theory~\cite{Son:2000xc}
predicts a gradual rotation of the condensates so that $\Sigma_{\bar\psi\psi}^2
+ \Sigma_\pi^2=1$ holds irrespective of $\mu_I$.

\section{Extrapolation in the pionic source}

The most crucial step in the analysis is the extrapolation $\lambda\to0$. As we
will see below, this is rather cumbersome, since the dependence on $\lambda$ 
around zero is pronounced for most of our observables. (The physical reason
behind such a strong dependence on $\lambda$ will be given in
Sec.~\ref{sec:valimp} below.) For low values of $T$, $\mu_I$ and $\lambda$, one
can use chiral perturbation theory~\cite{Splittorff:2002xn} to guide the
extrapolation, as was done in Ref.~\cite{Endrodi:2014lja}. For temperatures
close to $T_c(0)$, however, chiral perturbation theory is no longer valid. In
the direct vicinity of the phase boundary to the pion condensation phase one
expects the observables to be governed by the critical exponents of the
universality class, in this case supposedly $\mathrm{O}(2)$~\cite{Son:2000xc},
due to symmetry, associated with the transition. Away from criticality, however,
this is no longer true, so that other methods are needed for the extrapolation.

Note that in the $\lambda\to0$ limit the $\mathrm{U}_{\tau_3}(1)$ flavor
symmetry of the action (discussed in Sec.~\ref{sec:setup}) becomes exact and,
accordingly, the Goldstone mode associated with its spontaneous breaking exactly
massless. Strictly speaking, this implies that the thermodynamic limit
$V\to\infty$ should be performed prior to the $\lambda\to0$ extrapolation. We
address this subtle issue below in Sec.~\ref{sec:results} by comparing our
results on three different spatial volumes. 

\subsection{Naive extrapolations}

To perform the $\lambda$-extrapolation of the observables in a
model-independent way, we develop a spline extrapolation scheme that provides a
conservative systematic error on the $\lambda\to0$ value. The method involves a
fit of the data to a spline function defined in terms of a set of nodepoints,
complemented by a Monte-Carlo simulation of the nodepoints based on an action
weighing the fits according to the Akaike information
criterion~\cite{Borsanyi:2014jba}. The method is described in detail in
App.~\ref{app}.

\begin{figure}[ht!]
 \centering
 \includegraphics[width=.48\textwidth]{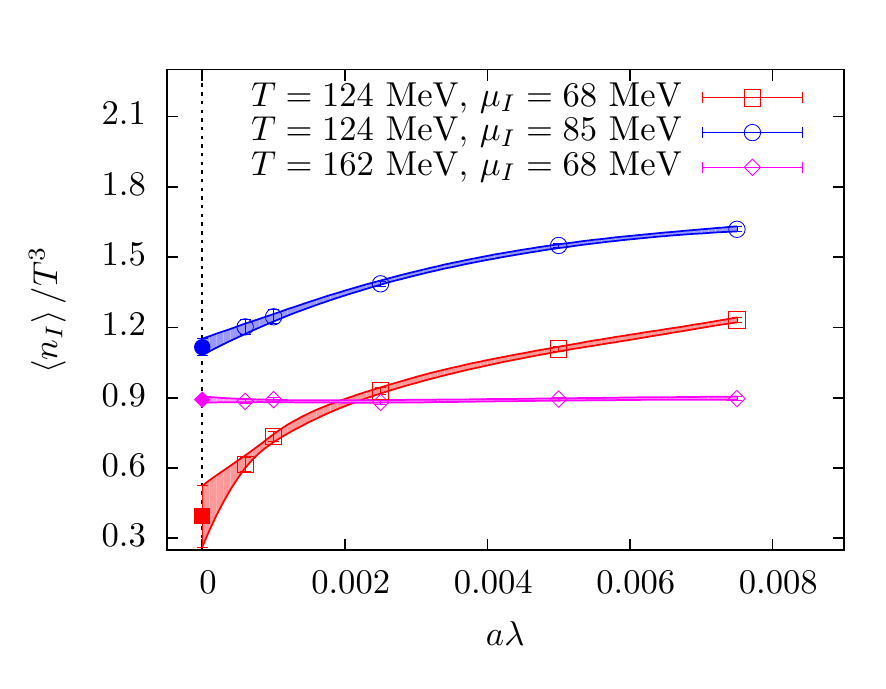}
 \caption{Results for the extrapolation of $n_I$ using the spline
 Monte-Carlo outlined in the appendix.}
 \label{fig:extrap}
\end{figure}

In Fig.~\ref{fig:extrap} we show typical examples of this extrapolation
scheme. The isospin density is plotted for three different parameter values on
$24^3\times6$ lattices with different $\lambda$-dependences. The plot shows
that the extrapolated values as well as the
associated uncertainties indeed give a reasonable and conservative
representation of the available data. In
particular, for a flat dataset the spline extrapolation is similarly flat with
a reasonably small uncertainty, which, in contrast to a simple linear
extrapolation, also takes a possible slight curvature into account. When the
extrapolation is steeper, the uncertainty of the extrapolated result increases
accordingly, signaling the enhanced impact of the last few points and the
associated loss in information on the extrapolated value.

\subsection{Valence quark improvement}
\label{sec:valimp}

\begin{figure}[t]
 \centering
 \begin{minipage}{.48\textwidth}
 \centering
 \includegraphics[width=\textwidth]{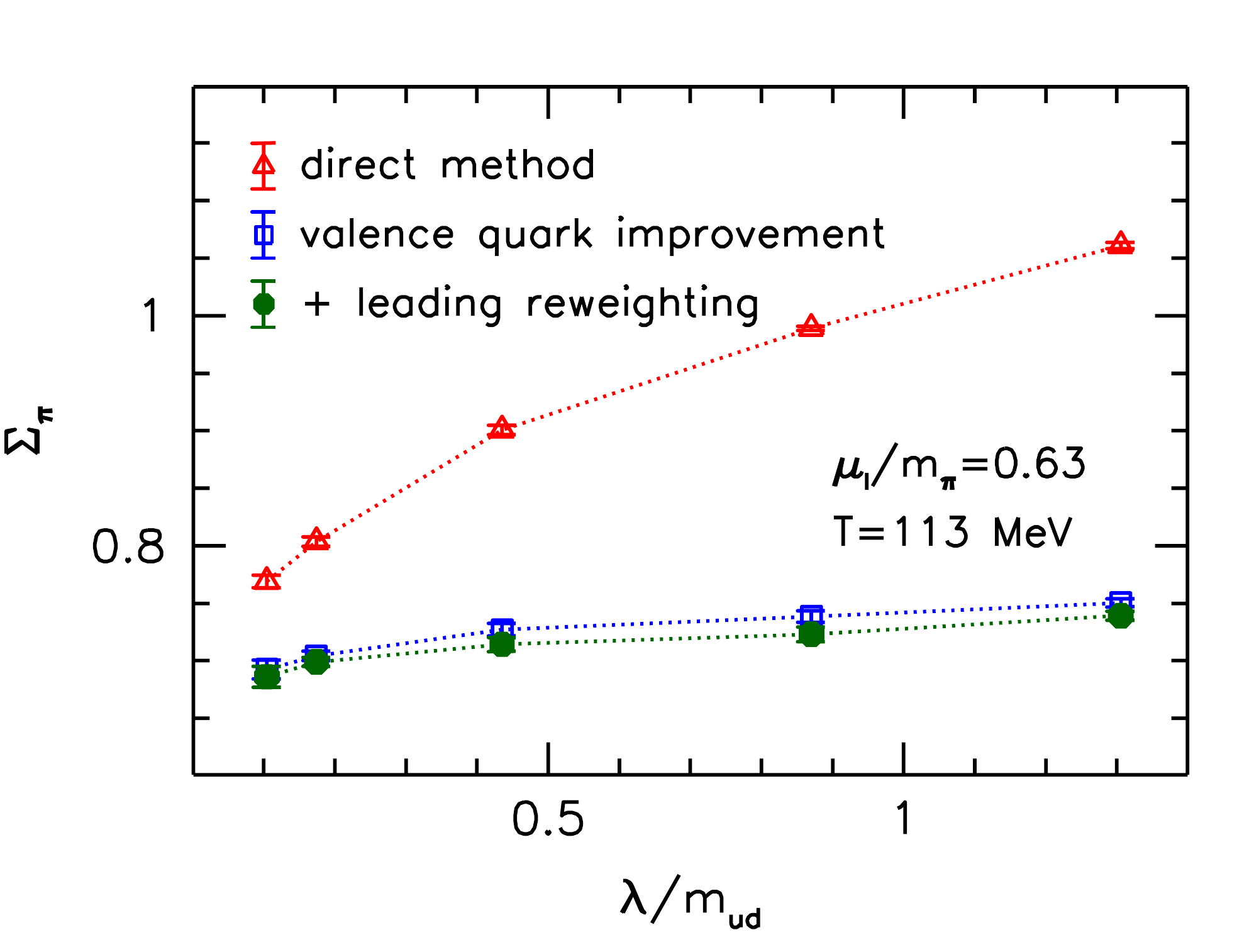}
 \end{minipage}
 \begin{minipage}{.48\textwidth}
 \centering
 \includegraphics[width=\textwidth]{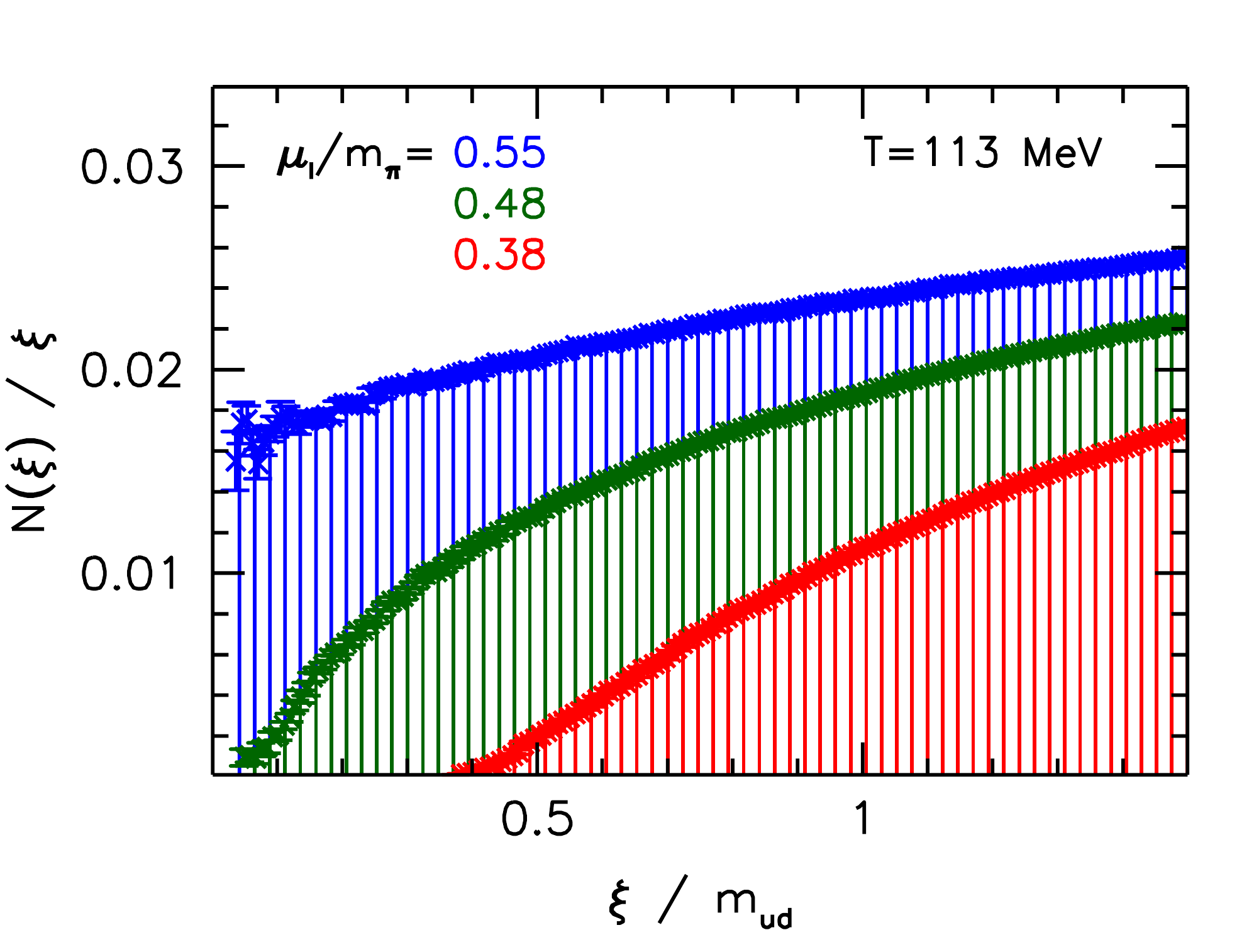}
 \end{minipage}
 \caption{
 {\bf Left:} Comparison of the $\lambda$-dependence of the different
 improved observables for the extrapolation of the pion condensate
 $\pi$. {\bf Right:} Integrated spectral density of the massive singular values 
 of the Dirac operator for various chemical potentials around
 the threshold value $m_\pi/2$ at low temperature $T=113\textmd{ MeV}$.}
 \label{fig:hist}
\end{figure}

The above method works well for the extrapolation and can be applied
successfully to all observables. Nevertheless, the accumulation of large
uncertainties for observables where the $\lambda$-extrapolation is particularly
steep is still a source for concern and a strong constraint on the accuracy
concerning the investigation of the phase diagram. This is particularly true for
the phase boundary to the pion condensation phase, which is defined by the onset
of a non-vanishing value of the pion condensate $\expv{\pi}$, an example is
shown in the left panel of Fig.~\ref{fig:hist}. To overcome this problem, we
introduce a novel approach, which uses the singular value representation of the
observables of Eq.~(\ref{eq:traces1}). While the basic idea can be applied to
all of these observables its application to the pion condensate takes a
particularly instructive form, which we will now discuss.

We begin with the singular value equation of the massive Dirac operator,
\be
|\Dp+m_\light|^2 \psi_n = \xi_n^2 \psi_n\,,
\ee
which involves the massive singular values $\xi_n>0$. 
In the basis spanned by $\psi_n$ we can rewrite the pion condensate of 
Eq.~(\ref{eq:traces1}) as
\be
\expv{\pi} = \frac{\lambda T}{2V} \expv{\sum_n \,(\xi_n^2+\lambda^2)^{-1}}
\xrightarrow{V\to\infty} \frac{\lambda }{2} \expv{\int \d \xi \,\rho(\xi)
(\xi^2+\lambda^2)^{-1}}
\xrightarrow{\lambda\to0} \frac{\pi }{4} \expv{\rho(0)}\,.
\label{eq:BC}
\ee
Here in the second step we considered the volume to be large enough so that 
the singular values become sufficiently dense and the sum can be replaced by 
an integral introducing the density $\rho(\xi)$ of the singular values (which
includes the normalization factor $T/V$). In the third step we performed the
$\lambda\to0$ limit, which leads to a representation of the $\delta$-function
and results in the density $\rho(0)$ around zero. This Banks-Casher-type
relation was first found in Ref.~\cite{Kanazawa:2011tt} for the massless case
$m_\light=0$. Here we have generalized it to the massive case so that the
singular values depend explicitly on $m_\light$. Eq.~(\ref{eq:BC}) reveals that
a nonzero pion condensate is equivalent to an accumulation of the near-zero
(massive) singular values of the Dirac operator. This also explains the strong
dependence of $\expv{\pi}$ on $\lambda$ -- the pionic source shifts the singular
value spectrum up by an amount $\lambda$ and thus impacts on the infrared
physics drastically.

To be more quantitative, in the right panel of Fig.~\ref{fig:hist} we plot the
integrated spectral density $N(\xi)=\int_0^\xi \d \xi' \rho(\xi')$ divided by
$\xi$, as measured on our $24^3\times6$ ensembles with $\lambda=0.001$ , for
three different isospin chemical potentials below, close to and above the onset
value $\mu_I=m_\pi/2$. Note that the $\xi\to0$ limit of this quantity gives
$\rho(0)$. The figure clearly shows that the density at zero vanishes in the
normal phase but develops a nonzero expectation value in the pion condensed
phase. The results for $\expv{\pi}$ obtained via $\expv{\rho(0)}$ on each
$\lambda>0$ ensemble are also shown in the left panel of Fig.~\ref{fig:hist}.
The plot indicates that this definition of $\pi$ leads to a drastic improvement
compared to the standard observable. Since this improved definition corresponds
to an explicit $\lambda=0$ substitution in the measured operator (in lattice
language: in the valence sector), we denote this type of improvement as
``valence quark improvement''. A similar improvement can be performed for the
other observables in Eq.~(\ref{eq:traces1}) as well -- this will be discussed in
a future publication.

\subsection{Leading-order reweighting}

The improvement of the operators $\O$ described above corrects for most of the
effects due to $\lambda>0$ in our observables. The remaining
$\lambda$-dependence originates from sea quarks, i.e.\ from the nonzero value of
$\lambda$ in the fermion determinant of the path integral measure. This can be
taken into account via the reweighting approach,
\be
\expv{\O}_{\lambda=0} = \frac{\expv{\O \,W(\lambda)}_{\lambda>0} }{
\expv{W(\lambda)}_{\lambda>0} }, \quad\quad W(\lambda)\equiv \frac{\det
\left[|\Dp+m_\light|^2\right]^{1/4}}{\det\left[
|\Dp+m_\light|^2+\lambda^2\right] ^{1/4}}\,,
\ee
where we used the relations in Eq.~(\ref{eq:positive}). Rewriting the logarithm
$\log W(\lambda)$ of the reweighting factor we obtain
\be
\log \left.\frac{\det
\left[|\Dp+m_\light|^2+\lambda^2-\lambda_w^2\right]^{1/4}}{\det\left[
|\Dp+m_\light|^2+\lambda^2\right] ^{1/4}}\right|_{\lambda_w=\lambda} =
\left[ -\lambda_w^2\frac{V}{2T} \frac{\pi}{\lambda} + \O(\lambda_w^4)
\right]_{\lambda_w=\lambda} 
= -\frac{\lambda V}{2T}\pi + \O(\lambda^4)\,,
\ee
where we replaced the expression with its Taylor-expansion in $\lambda_w^2$
around $\lambda_w=0$ and compared to the pion condensate~(\ref{eq:traces1}).
Thus we conclude that the reweighting of an observable, to leading order in
$\lambda$, involves the exponential of the pion condensate times the
four-volume. Since the pion condensate is anyway measured for $\expv{\pi}$, this
improvement comes with no extra costs. The inclusion of the reweighting factors
reduces the dependence of the pion condensate on $\lambda$ further, as visible
in the left panel of Fig.~\ref{fig:hist}.

\section{Results for the phase diagram}
\label{sec:results}

We will now discuss the results concerning the phase diagram. The focus of the
present contribution is on lattices with a temporal extent $N_t=6$, but in the
next section we will also include results from $N_t=8,10$ and 12 lattices. We
have found that the introduction of the pionic source is necessary for the
stability of the simulations throughout the phase diagram -- even when we are
not within the phase where pions condense. The extrapolation of the results 
to $\lambda=0$ is performed using the methods discussed above.

\subsection{Phase boundary to the pion condensation phase}
\label{sec:pion}

\begin{figure}[ht!]
 \centering
 \begin{minipage}{.48\textwidth}
 \centering
 \includegraphics[width=\textwidth]{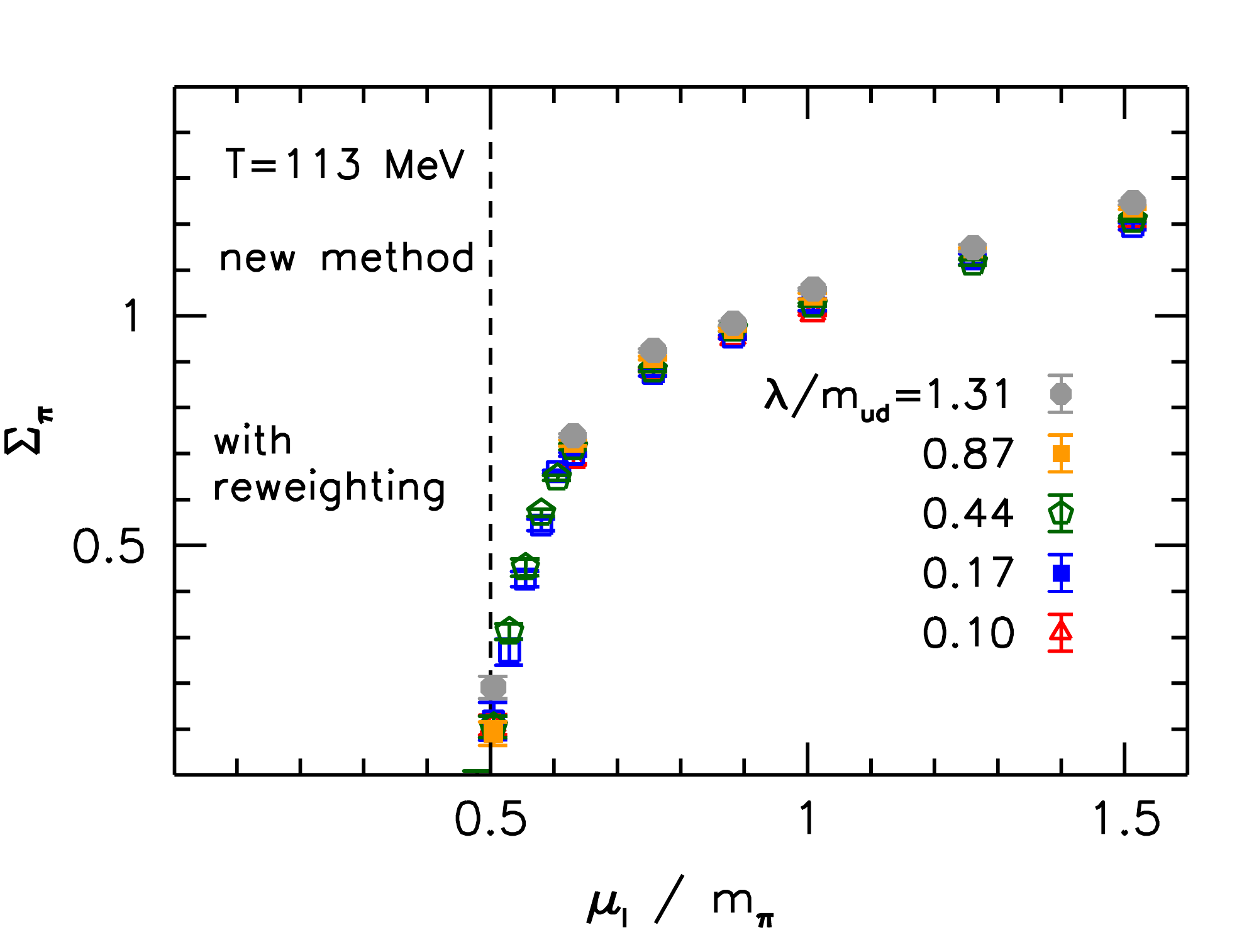}
 \end{minipage}
 \begin{minipage}{.48\textwidth}
 \centering
 \includegraphics[width=\textwidth]{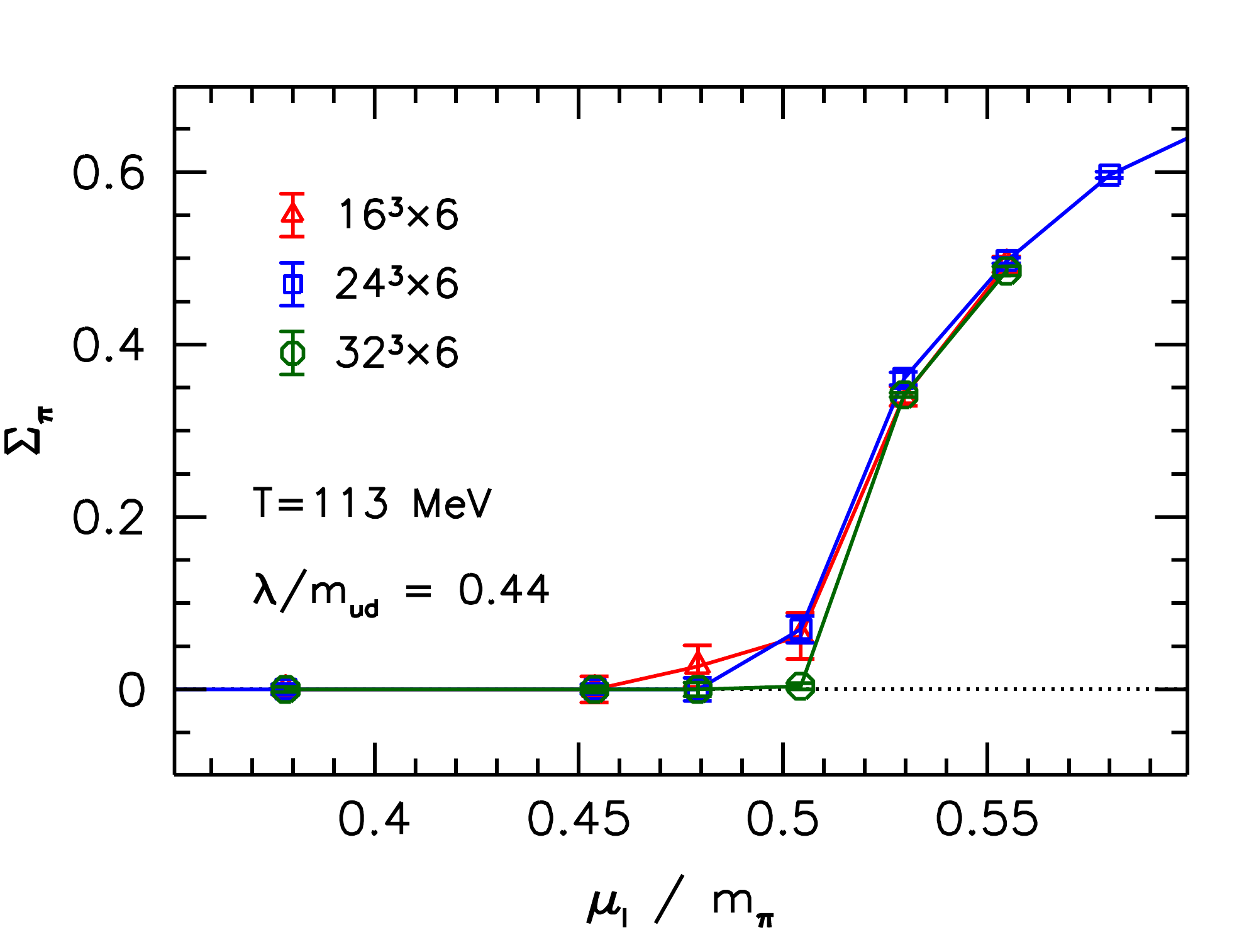}
 \end{minipage}
 \caption{
 {\bf Left:} Results for the improved and renormalized pion condensate on the
 $24^3\times6$ lattice at $T=113$ MeV versus $\mu_I$. The vertical line
 indicates the boundary to the pion condensation phase at $T=0$,
 $\mu_I/m_\pi=0.5$. {\bf Right:} The volume-dependence of the same
 observable at fixed temperature and pionic source.}
  \label{fig:pion}
\end{figure}

The phase boundary to the pion condensation phase is defined by the point where
the system develops a non-zero pion condensate. We show the typical behavior of
the improved and renormalized pion condensate, including also the leading order
reweighting factor, for different values of $\lambda$ in the left panel of
Fig.~\ref{fig:pion}. As can be seen from the plot, the phase boundary is clearly
visible even for finite values of $\lambda$ and the results from different
$\lambda$-values, owing to the use of the improved operator, fall on top of each
other. In the right panel of the figure we show the volume-dependence of
$\Sigma_{\pi}$ using three different spatial lattice sizes. This reveals the
typical sharpening around the critical chemical potential, suggesting that a
real phase transition takes place in the $V\to\infty$ limit.

\begin{figure}[ht!]
 \centering
 \includegraphics[width=.48\textwidth]{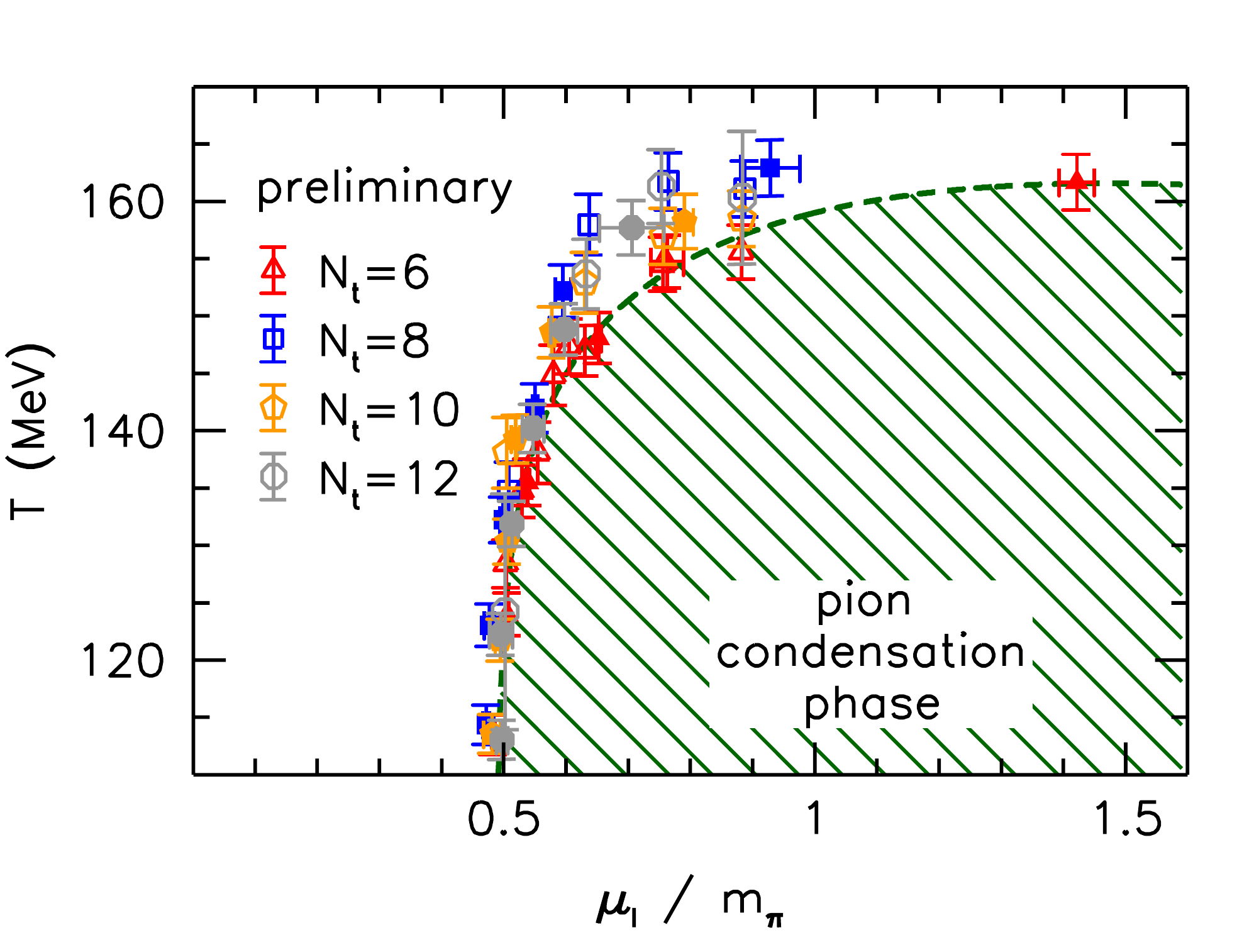}
 \caption{\label{fig:pion2}
  Results for the boundary of the pion
 condensation phase in the $(T,\mu_I)$ parameter plane. Open symbols are
 obtained from scans in the temperature, filled symbols from scans
 in $\mu_I$. Besides statistical errors, the latter are also
 subject to an intrinsic  uncertainty for $T$ originating from
 the lattice scale.
 }
\end{figure}

In Fig.~\ref{fig:pion2} we show the associated results for the phase boundary in
the $(T,\mu_I)$ parameter plane for $N_t=6,8,10$ and 12. The plot indicates that
the boundary of the pion condensation phase basically follows the vertical
$\mu_I/m_\pi=0.5$ line up to a temperature of about $T\approx130$ MeV,
independent of the value of $N_t$. More surprising is the strong flattening of
the phase boundary at $T\approx160$ MeV, which is in qualitative agreement with
the findings
from~\cite{Kogut:2002tm,Kogut:2002zg,Kogut:2004zg,deForcrand:2007uz}, but in
contrast to the expectations from chiral perturbation theory~\cite{Son:2000xc}.
Above $T\approx130$ MeV the phase boundary also becomes more sensible to lattice
artefacts, indicated by the larger spread of the points from different $N_t$
values.

\subsection{Crossover line at small density}

Next, we investigate the behavior of the crossover line in the
$(T,\mu_I)$-plane, starting from the well-known crossover at the physical point
at $\mu_I=0$. The main observable associated with the crossover, i.e., with the
restoration of chiral symmetry, is the renormalized chiral condensate of
(Eq.~\ref{eq:renobs}). The pseudocritical temperature of the crossover can be
defined, for instance, by the inflection point of the condensate, see, e.g.,
Ref.~\cite{Borsanyi:2010bp}. Using this definition and the same action as in our
study (for $\mu_I=0$), the crossover temperature in the continuum limit was
determined to be $T_c(0)=155(3)(3)$~\cite{Borsanyi:2010bp}. Here we will use a
slightly different definition and define the crossover temperature to be the
temperature where the renormalized chiral condensate acquires its continuum
value at $T_c(0)$. Following Ref.~\cite{Borsanyi:2010bp} the value is given by
$\left.\Sigma_{\bar\psi\psi}\right|_{T_c}\approx-0.550$ in our normalization.
This definition for $T_c(\mu_I)$ is adequate as long as we are in the Silver
Blaze region (where the $T\to0$ limit of the condensate is independent of $\mu_I$) 
and should be compared to the results from other definitions eventually.

\begin{figure}[ht!]
\centering
\begin{minipage}{.48\textwidth}
\centering
\includegraphics[width=\textwidth]{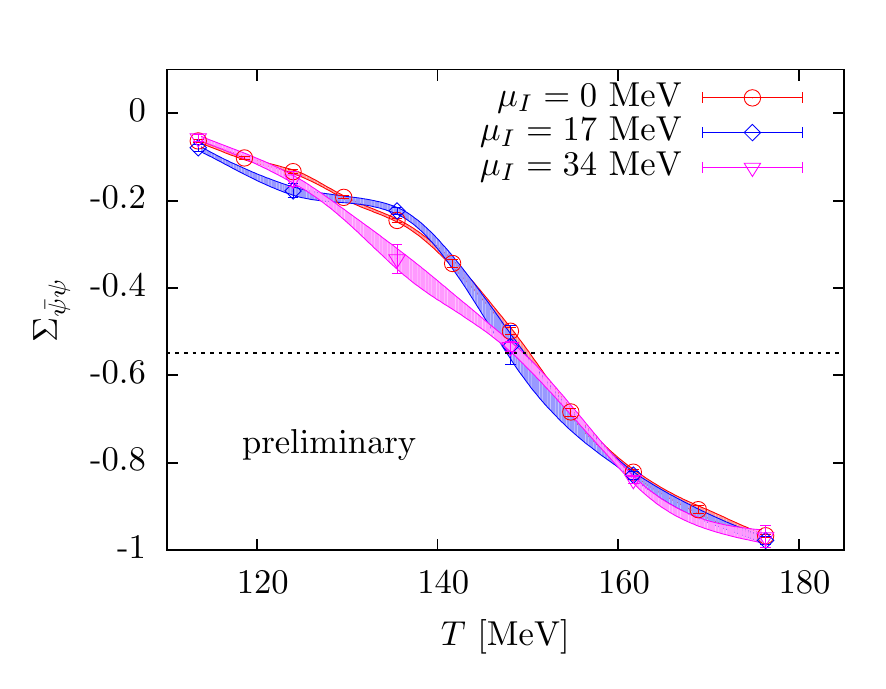}
\end{minipage}
\begin{minipage}{.48\textwidth}
\centering
\includegraphics[width=\textwidth]{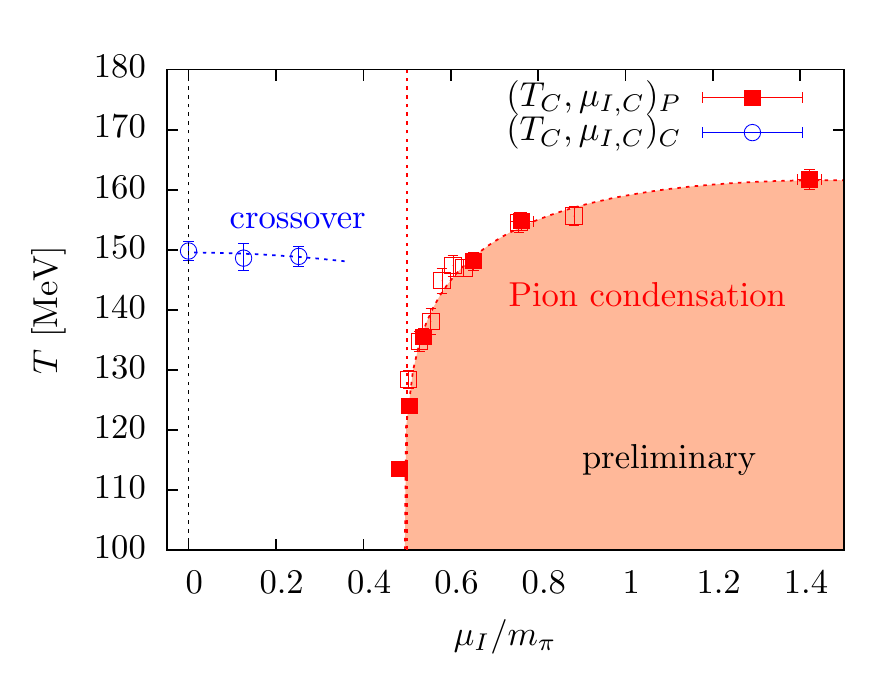}
\end{minipage}
\caption{
 {\bf Left:} Results for $\Sigma_{\bar\psi\psi}$ vs. the temperature for
 different values of $\mu_I$.
 The colored areas are the results from a cubic spline interpolation and the
 dashed horizontal line indicates $\left.\Sigma_{\bar\psi\psi}\right|_{T_c}$
 (see text). {\bf Right:} Phase diagram for the $24^3\times6$ lattice. The red
 points are the results for the phase boundary to the pion condensation phase,
 $(T_c,\mu_{I,c})_P$, and the blue points the ones
 for the crossover line, $(T_c,\mu_{I,c})_C$.}
 \label{fig:cross}
\end{figure}

In the left panel of Fig.~\ref{fig:cross} we show the results for the
renormalized chiral condensate versus the temperature for different values of
$\mu_I$. The horizontal line indicates the value
$\left.\Sigma_{\bar\psi\psi}\right|_{T_c}$ and the colored bands result from
cubic spline interpolations of the data points. For $\mu_I=0$ the result for
$T_c(0)=150(3)(3)$ (where the second uncertainty is due to scale setting) is
slightly smaller than the continuum result. This can be attributed to lattice
artifacts which can still be sizable at $N_t=6$. The plot shows that the result
for $T_c$ does not change significantly up to $\mu_I=34$ MeV.

In the right panel of Fig.~\ref{fig:cross} we show the resulting phase diagram
for the $24^3\times6$ lattice. At the physical pion mass, the crossover
temperature at $\mu_I=0$ appears to lie somewhat below the temperature
associated with the melting of the pion condensate (i.e.\ the upper boundary of
the pion condensation phase). However, notice that the chiral restoration
transition is a broad crossover, whereas pion condensation sets in via a real
phase transition. The nature of the latter transition will be the subject of a
forthcoming publication. Hints for a critical endpoint, where the transition
turns first order, were obtained on small lattices and heavier-than-physical 
quark masses in Refs.~\cite{Kogut:2002zg,Kogut:2004zg}.

A slight downwards trend in $T_c(\mu_I)$ is observed as the chemical potential
increases. The dashed curve corresponds to a quadratic fit to the crossover
temperatures. Translating the included coefficient $\kappa$ into the
normalization from~\cite{Endrodi:2011gv} and using that, in the present setup,
$3\mu_I$ compares to $\mu_B$, we obtain $\kappa\approx0.015$, which is of the
same order of magnitude as the analogous results for nonzero baryon chemical
potentials, from~\cite{Bellwied:2015lba}.

\section{A test for Taylor expansion}

One of the long-standing challenges in lattice QCD is the complex action
problem, mentioned in the introduction, which hinders direct simulations at
non-zero baryon chemical potential $\mu_B$. A standard technique to circumvent
this problem is the Taylor expansion method, which involves working with
derivatives with respect to $\mu_B$ evaluated at $\mu_B=0$. One of the major
drawbacks of this approach is the {\it a priori} unknown range of applicability 
when working at a fixed order of the expansion. Since the Taylor expansion can
be defined with respect to any of the chemical potentials in
Eq.~(\ref{eq:chpot}), our results at finite isospin chemical potential
provide an ideal testbed for the method.

For the comparison of our direct results to the Taylor method we will focus on
the isospin density $n_I$ from Eq.~(\ref{eq:pbpdef}). Its expansion in terms of
$\mu_I$ can be written in the form
\be
\frac{\left\langle n_I \right\rangle}{T^3} = c_2 \Big(\frac{\mu_I}{T}\Big) +
\frac{c_4}{6} \Big(\frac{\mu_I}{T}\Big)^3 \,.
\ee
Here $c_2$ and $c_4$ are Taylor coefficients, which, using the expression
for the QCD pressure
\be
\frac{p}{T^4} = \frac{1}{VT^3} \log \Z \,,
\ee
can be written as
\be
c_2=2 \Big[\partial_u^2\Big(\frac{p}{T^4}\Big) - \partial_u \partial_d
\Big(\frac{p}{T^4}\Big) \Big] \quad \text{and} \quad  c_4=2\Big[ \partial_u^4
\Big(\frac{p}{T^4}\Big) - 4 \partial_u^3 \partial_d \Big(\frac{p}{T^4}\Big) + 3
\partial_u^2 \partial_d^2 \Big(\frac{p}{T^4}\Big) \Big] \,,
\ee
where $\partial_f$ is the derivative with respect to the chemical potential to
temperature ratio $\mu_f/T$. We compare to the Taylor expansion coefficients
determined in Ref.~\cite{Borsanyi:2011sw}, which uses the same action as we do
in the present study. To arrive at the coefficients corresponding to the
temperatures of our simulations we have performed a cubic spline interpolation
of $c_2$ and of $c_4$.

\begin{figure}[ht!]
\centering
\begin{minipage}{.48\textwidth}
\centering
\includegraphics[width=\textwidth]{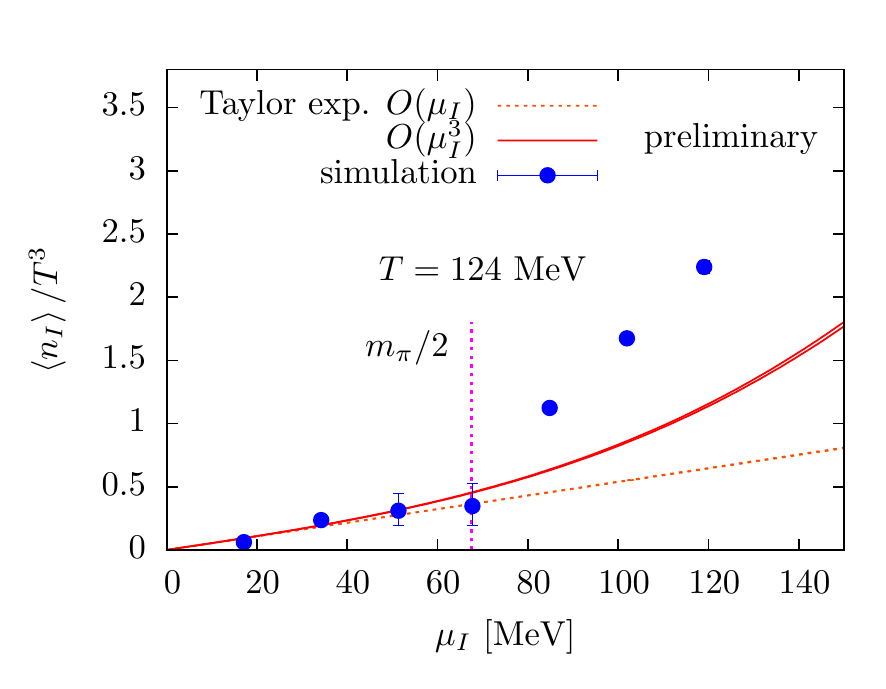}
\end{minipage}
\begin{minipage}{.48\textwidth}
\centering
\includegraphics[width=\textwidth]{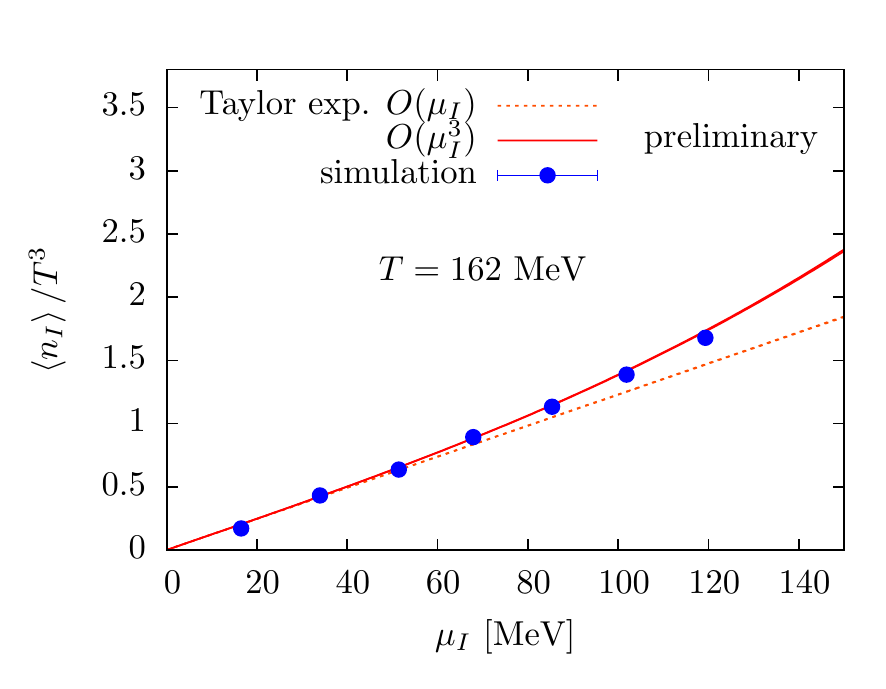}
\end{minipage}
\caption{
 Comparison of the results for the isospin density from $24^3\times6$ lattices
 for temperatures 124 (left) and 162 MeV (right) and the results from Taylor
 expansion around $\mu_I=0$ to $\O(\mu_I)$ and $\O(\mu_I^3)$. The dashed vertical
 line in the left panel indicates the phase boundary to the pion condensation
 phase.}
 \label{fig:taylor}
\end{figure}

The comparison for the $N_t=6$ lattices is shown in Fig.~\ref{fig:taylor} for
two temperatures of 124 (left) and 162 MeV (right).\footnote{We mention that our
$N_s=24$ ensembles are compared here to the $N_s=18$ results of
Ref.~\cite{Borsanyi:2011sw}. Nevertheless, we have explicitly checked by
comparing results for the coefficients obtained on different
volumes~\cite{Borsanyi:2011sw}, that finite size effects on the $c_i$ are
negligible.} Based on the last section, the former temperature is below $T_c(0)$
and below the temperature associated with the melting of the pion condensate, so
that we enter the pion condensation phase at $\mu_I/m_\pi\approx0.5$, indicated
by the vertical line in Fig.~\ref{fig:taylor} (left). The latter temperature is
above $T_c(0)$ and above the phase boundary of the pion condensation phase (at
least for the values of $\mu_I$ considered here), so that we remain in the
normal phase.

For $T=124$ MeV the data agrees very well with the Taylor expansion up to the
point where we enter the pion condensation phase. At this point the ground state
of the system changes drastically, so that, naturally, one expects the Taylor
expansion method to break down. This is indeed visible in the plot: up to this
value of $\mu_I$ the Taylor expansion curves to $\O(\mu_I)$ and $\O(\mu_I^3)$
are still too close to be distinguished with the present accuracy of the data.
For $T=162$ MeV the agreement between the data and the Taylor expansion to
$\O(\mu_I^3)$ persists for all values of $\mu_I$ considered at present (perhaps
with the exception of the last point). Starting from $\mu_I/m_\pi\approx0.5$ the
term of order $\O(\mu_I^3)$ becomes important, so that the Taylor expansions to
$\O(\mu_I)$ and $\O(\mu_I^3)$ can be distinguished. It would be interesting to
generate data at larger values of $\mu_I$ to see how long the agreement between
the data and the expansion to $\O(\mu_I^3)$ remains and we will investigate this
in a future publication.

\section{Conclusions}

In this contribution we have presented first results from our investigation of
the phase diagram of QCD in the presence of a finite isospin chemical
potential, using physical quark masses and an improved staggered action.
We have introduced, for the first time in this context, the massive singular
value representation of the partition function and of the relevant observables.
This representation lead us to a Banks-Casher-type relation that connects the
pion condensate to the singular value density at zero. Employing this relation
was found to drastically improve the extrapolation $\lambda\to0$ in the pionic
source -- this auxiliary parameter $\lambda>0$ is necessary in the simulations
to have the infrared behavior of the system under control. Besides this valence
improvement, we included reweighting factors to take into account the
leading-order effect of the pionic source for the sea quarks as well. Finally,
we worked out a spline Monte-Carlo scheme for performing additional
extrapolations in $\lambda$ in a model-independent manner.

At high values of $\mu_I$ the system was found to be in the pion condensed
phase, separated from the normal phase -- as our preliminary finite-volume
analysis suggests -- by a real phase transition. We have mapped out this phase
boundary and found it, on the one hand, to be almost $T$-independent for
temperatures below the $\mu_I=0$ chiral crossover temperature $T_c(0)\approx
155\textmd{ MeV}$. On the other hand, for higher temperatures the phase boundary
appears to flatten out so that no pion condensate forms above $\sim160\textmd{
MeV}$, at least for the chemical potentials that we investigated here. A
simplistic interpretation of these findings is that in the chirally
restored/deconfined phase pions do not exist anymore, thus the formation of a
pion condensate becomes strongly suppressed in this region.

For low chemical potentials, we calculated the dependence of the 
chiral/deconfinement cross\-over temperature on $\mu_I$ and made a rough
estimate on the curvature of the transition line. In addition, we performed a
quantitative comparison of our direct results to the leading- and the
next-to-leading-order Taylor expansion in the isospin chemical potential at
$\mu_I=0$. Currently all results for the isospin density show good agreement
with the Taylor expansion to $\O(\mu_I^3)$, as long as the boundary of the pion
condensation phase is not crossed. The contributions from $\O(\mu_I^3)$ start to
become important at $\mu_I/T\approx0.5$ for $T=162$ MeV so that the expansions
to $\O(\mu_I)$ and $\O(\mu_I^3)$ can be distinguished. It will be interesting to
simulate at larger values of $\mu_I$ to check where terms of $\O(\mu_I^5)$
become non-negligible.

To extend the present study, we plan to apply the novel techniques developed
here to map out the complete phase diagram concentrating, in particular, on the
meeting point of the two different transition lines and on the high-$\mu_I$
region.\\

\noindent
{\bf Acknowledgments}\; 
This research was funded by the DFG (Emmy Noether Programme EN 1064/2-1 and
SFB/TRR 55). The majority of the simulations was performed on the GPU cluster 
of the Institute for theoretical Physics at the University of Regensburg.
B.\ B.\ acknowledges support from the Frankfurter
F\"orderverein f\"ur Physikalische Grundlagenforschung.
The authors thank Szabolcs Bors\'anyi for useful correspondence and 
for providing the data for the Taylor expansion coefficients.

\appendix

\section{A model independent spline extrapolation}
\label{app}

A model-independent extrapolation should take into account all smooth functions
that go through the available data points. A possible way to represent this
function space is by means of polynomials of degree $n$, crafted together at a
set of grid points, i.e.\ a spline. A particular version of spline
interpolation is spline fitting (see, e.g., Ref.~\cite{Endrodi:2010ai}), where
the grid points are placed around and between the datapoints, so that the free
parameters of the spline are determined by a fit to the data. A spline, for
which the outermost grid points lie outside of the dataset constitutes an
extrapolation. Typical spline fits involve the use of so-called natural boundary
conditions at these outermost grid points, setting the second and higher
derivatives there to zero. For the most general curve in case of an
extrapolation, these higher derivatives should also be treated as free
parameters.

The remaining issue with the resulting extrapolation is the dependence of the
result on the number and position of the grid points. To ensure
model-independence, we need to average over all possible grid point
configurations. However, different sets of grid points might not allow for
equivalently good descriptions of the data. Taking this into account turns
the average over grid points into a weighted average. There are several possible
weight factors that one can use. We will write the weight factor in the form
$w=\exp(-S_{\rm spl})$, where we have introduced the ``action'' $S_{\rm spl}$.
Then the average over all possible spline configurations of a quantity $A$ can
be written as
\be
\big\llangle A \big\rrangle = \sum_{N_G} \int d^{N_x(N_G)}x
\,\, A(\vec{x}) \, \exp\big(-S_{\rm spl}(\vec{x},N_G)\big) \,.
\ee
Here the sum runs over the possible number $N_G$ of spline grid points, we have
denoted the set of grid points by $\vec{x}$ and $N_x(N_G)$ is the number of grid
points that can be varied. Note that the positions of the data points can also
be restricted to certain areas. In particular, in our setup we use a rectangular
grid and demand that in each direction at least one measurement is included in
each interval between two grid points. The weighted sum can be performed
efficiently using Monte-Carlo methods~\cite{Szabolcs}, e.g. via a Metropolis
algorithm.

A possible choice for the action is to use the Akaike information
criterion~\cite{Borsanyi:2014jba},
\be
\label{eq:Saic}
S_{\rm AIC} = 2 N_P + \chi^2 \,,
\ee
where $N_P$ is the number of parameters of the fit. An alternative choice
is to use the goodness of the fit, leading to the action
\be
\label{eq:Sgod}
S_{\rm GOOD} = P(\chi^2,N_{\rm dof}) - 1 \,, \quad \text{where} \quad
P(\chi^2,N_{\rm dof}) = \frac{\gamma(\chi^2/2,N_{\rm dof}/2)}{\Gamma(N_{\rm
dof}/2)}
\ee
is the cumulative $\chi^2$-distribution function, $N_{\rm dof}$ the number of
degrees of freedom and $\gamma$ the lower incomplete gamma-function. We have
tested these definitions and found that the results of the
$\lambda$-extrapolations from the Akaike information criterion were more robust
with respect to oscillatory spline solutions (see~\cite{Endrodi:2010ai} for a
detailed discussion), most probably due to the explicit punishment for a large
number of fit parameters.

In the present setting, we perform a one-dimensional spline extrapolation
in $\lambda$ for each set of bare parameters ($\beta$, $\mu_I$). We use cubic
splines and set the second derivative of the curve at the nodepoint for the
highest $\lambda$ to zero and leave the second derivative at $x_1=0$ a free
parameter. For the spline Monte-Carlo we employ the action from
Eq.~(\ref{eq:Saic}) and use 1000 measurements, separated by 20 updates and
preceded by 1000 thermalization updates. One more compromise we had to make in
order to obtain stable results was to set the lowest nodepoints so that at least
two datapoints lie between $x_1=0$ and $x_2$.

To estimate the uncertainties from the spline Monte-Carlo we have performed the
spline fits for each bootstrapped sample, using the action associated with the
fit to the central value in the update steps. The resulting uncertainty for the
extrapolated values consists of two uncertainties: (i) the statistical
uncertainty, obtained from the bootstrap samples containing the average over all
spline fits for this sample; (ii) the Monte-Carlo uncertainty on the central
value, obtained in the standard way from the individual fits to the central
value. These two types of errors have been conservatively added in quadrature.

\providecommand{\href}[2]{#2}\begingroup\raggedright\endgroup

\end{document}